\documentclass[prd,showkeys,floatfix,twocolumn,amsmath,amssymb,floatfix]{revtex4}
\usepackage{graphicx,color,dcolumn,booktabs,bm}
\usepackage{longtable,lscape}
\usepackage{amssymb}
\usepackage{indentfirst}
\usepackage{epsfig}
\usepackage{feynmf}
\usepackage{epstopdf}
\usepackage{slashed}
\usepackage{cases}
\usepackage{multirow}
\usepackage[colorlinks,citecolor=blue,anchorcolor=red,menucolor=red,linkcolor=red,filecolor=red,runcolor=red,urlcolor=blue,frenchlinks=red]{hyperref}

\allowdisplaybreaks[3]

\begin{document}
\title{Light tetraquark states with the exotic quantum number $J^{PC} = 3^{-+}$}
%

\author{Niu Su$^{1,2}$}
\author{Rui-Rui Dong$^{1,2}$}
\author{Hua-Xing Chen$^1$}
\email{hxchen@seu.edu.cn}
\author{Wei Chen$^3$}
\email{chenwei29@mail.sysu.edu.cn}
\author{Er-Liang Cui$^4$}
\email{erliang.cui@nwafu.edu.cn}

\affiliation{
$^1$School of Physics, Southeast University, Nanjing 210094, China\\
$^2$School of Physics, Beihang University, Beijing 100191, China\\
$^3$School of Physics, Sun Yat-Sen University, Guangzhou 510275, China\\
$^4$College of Science, Northwest A\&F University, Yangling 712100, China
}

\begin{abstract}
We apply the method of QCD sum rules to study the $s q \bar s \bar q$ tetraquark states with the exotic quantum number $J^{PC} = 3^{-+}$, and extract mass of the lowest-lying state to be $2.33^{+0.19}_{-0.16}$~GeV. To construct the relevant tetraquark currents we need to explicitly add the covariant derivative operator. Our systematical analysis on their relevant interpolating currents indicates that: a) this state well decays into the $P$-wave $\rho\phi/\omega\phi$ channel but not into the $\rho f_2(1525)/\omega f_2(1525)/\phi f_2(1270)$ channels, and b) it well decays into the $K^*(892) \bar K_2^*(1430)$ channel but not into the $P$-wave $K^*(892) \bar K^*(892)$ channel.
\end{abstract}
\maketitle
\pagenumbering{arabic}
%
%
%
\section{Introduction}\label{sec:intro}
%

There have been many candidates of exotic hadrons observed in particle experiments, which can not be well explained in the traditional quark model~\cite{pdg,Liu:2019zoy,Lebed:2016hpi,Esposito:2016noz,Guo:2017jvc,Ali:2017jda,Olsen:2017bmm,Karliner:2017qhf,Brambilla:2019esw,Guo:2019twa}. Many of them still have ``traditional'' quantum numbers that traditional $\bar q q$ mesons and $qqq$ baryons can also have. This makes them not so easy to be clearly identified as exotic hadrons. However, there exist some ``exotic'' quantum numbers that traditional hadrons can not have, such as the spin-parity quantum numbers $J^{PC} = 0^{--}$, $0^{+-}$, $1^{-+}$, $2^{+-}$, and $3^{-+}$, etc. These ``exotic'' quantum numbers are of particular interests, because the hadrons with such quantum numbers can not be explained as traditional hadrons any more. Such hadrons are definitely exotic hadrons, whose possible interpretations are tetraquark states~\cite{Chen:2008qw,Chen:2008ne,Jiao:2009ra,Huang:2016rro,LEE:2020eif,Du:2012pn,Fu:2018ngx}, hybrid states~\cite{Meyer:2015eta,Chetyrkin:2000tj,Zhang:2013rya,Huang:2014hya,Huang:2016upt,Ho:2018cat}, and glueballs~\cite{Qiao:2014vva,Qiao:2015iea,Pimikov:2017bkk}, etc. Note that these different exotic structures may mix together, and there would exist various possibilities whenever there found a state in experiment with some exotic quantum number.

Among the above exotic quantum numbers, the hybrid states of $J^{PC} = 1^{-+}$ have been extensively studied, since they are predicted to be the lightest hybrid states~\cite{Meyer:2015eta} and there are some experimental evidences on their existence~\cite{Thompson:1997bs,Abele:1999tf,Adams:2006sa}. The light tetraquark states of $J^{PC} = 1^{-+}$ have also been studied in Refs.~\cite{Chen:2008qw,Chen:2008ne} using the method of QCD sum rules, and their masses and possible decay channels were predicted there for both isospin-0 and isospin-1 states. Later the same QCD sum rule method was applied to extensively study light tetraquark states of $J^{PC} = 0^{--}$ in Refs.~\cite{Jiao:2009ra,Huang:2016rro,LEE:2020eif}, and those of $J^{PC} = 0^{+-}$ in Refs.~\cite{Du:2012pn,Fu:2018ngx}.

In this paper we shall investigate the exotic quantum number $J^{PC} = 3^{-+}$, and the other one $J^{PC} = 2^{+-}$ will be studied in future. We shall investigate the light $q s \bar q \bar s$ ($q=up/down$ and $s=strange$) tetraquark states with such a quantum number. They may exist in the energy region around 2.0~GeV. With a large amount of $J/\psi$ sample, the BESIII Collaboration are carefully examining the physics happening in this energy region~\cite{Bai:2003sw,Ablikim:2005um,BESIII:2010krt,Ablikim:2010au,Ablikim:2016hlu,Ablikim:2019zyw,Ablikim:2020pgw}. So do the Belle-II~\cite{Kou:2018nap} and GlueX~\cite{Austregesilo:2018mno} experiments. Hence, these states are potential exotic hadrons to be observed in future experiments. There has not been any theoretical study directly on this subject. In Ref.~\cite{Zhu:2013sca} the authors used the one-boson-exchange model to study the $D^* \bar D_2^*$ molecular state of $J^{PC} = 3^{-+}$. They found that the isoscalar ($I=0$) state has the most attractive potential, suggesting that this $D^* \bar D_2^*$ molecular state of $J^{PC} = 3^{-+}$ may exist, and the $K^*(892) \bar K_2^*(1430)$ molecular state of $J^{PC} = 3^{-+}$ might also exist. Besides, there was a Lattice QCD study on the $3^{-+}$ glueball, but this was done forty years ago~\cite{Shen:1985px}.

In this paper we shall investigate the $q s \bar q \bar s$ tetraquark state with the exotic quantum number $J^{PC} = 3^{-+}$ using the method of QCD sum rules. Recently, we have applied this method to study the $s s \bar s \bar s$ tetraquark states of $J^{PC} = 1^{\pm-}$ in Refs.~\cite{Chen:2008ej,Chen:2018kuu,Cui:2019roq}. In the present study we shall improve it by explicitly adding the covariant derivative operator in order to construct the $q s \bar q \bar s$ tetraquark currents of $J^{PC} = 3^{-+}$. This will be detailedly discussed in the next section.

This paper is organized as follows. In Sec.~\ref{sec:current}, we systematically construct the $q s \bar q \bar s$ tetraquark currents with the exotic quantum number $J^{PC} = 3^{-+}$. Then we use them to perform QCD sum rule analyses in Sec.~\ref{sec:sumrule}, and perform numerical analyses in Sec.~\ref{sec:numerical}. The results are summarized and discussed in Sec.~\ref{sec:summary}, where we discuss their special decay behavior.

%
\section{Interpolating Currents}\label{sec:current}
%

In this section we construct the $q s \bar q \bar s$ ($q=up/down$ and $s=strange$) tetraquark currents with the exotic quantum number $J^{PC} = 3^{-+}$. This quantum number is exotic, and can not be simply composed by using one quark and one antiquark. Moreover, we can not use only two quarks and two antiquarks without derivatives, and two quarks and two antiquarks together with at least one derivative are necessary to reach such a quantum number.

Besides, the $s s \bar s \bar s$ tetraquark currents of $J^{PC} = 3^{-+}$ can not be constructed using two quarks and two antiquarks with just one derivative; in the present study we shall not investigate the $q q \bar q \bar q$ tetraquark currents, since the widths of the $q s \bar q \bar s$ tetraquark states (if exist) are probably narrower, making them easier of being observed.

First let us consider the diquark-antidiquark $[q s] [\bar q \bar s]$ construction. In principle, the derivative can be either inside the diquark/antidiquark field or between the diquark and antidiquark fields, {\it i.e.},
\begin{eqnarray}
\eta &=& \big[q_a^T C \Gamma_1 {\overset{\leftrightarrow}{D}}_\alpha s_b \big] (\bar{q}_c \Gamma_2 C \bar{s}_d^T) \, ,
\label{eq:derivative1}
\\
\eta^\prime &=& (q_a^T C \Gamma_1 s_b) \big[\bar{q}_c \Gamma_2 C {\overset{\leftrightarrow}{D}}_\alpha \bar{s}_d^T\big] \, ,
\label{eq:derivative2}
\\
\eta^{\prime\prime} &=& \big[(q_a^T C \Gamma_3 s_b){\overset{\leftrightarrow}{D}}_\alpha(\bar{q}_c \Gamma_4 C \bar{s}_d^T)\big] \, ,
\label{eq:derivative3}
\end{eqnarray}
where $\big[ X {\overset{\leftrightarrow}{D}}_\alpha Y \big] = X [D_\alpha Y] - [D_\alpha X] Y$, with the covariant derivative $D_\alpha = \partial_\alpha + i g_s A_\alpha$; $a \cdots d$ are color indices, and the sum over repeated indices is taken; $\Gamma_{1,2,3,4}$ are Dirac matrices. However, we find that only the former construction can reach the quantum number $J^{PC} = 3^{-+}$.

Altogether we find six non-vanishing diquark-antidiquark currents of $J^{PC} = 3^{-+}$:
%
\begin{widetext}
\begin{eqnarray}
\eta^{1}_{\alpha_1\alpha_2\alpha_3} &=&
\epsilon^{abe} \epsilon^{cde} \times \mathcal{S} \Big\{
\big[q_a^T C \gamma_{\alpha_1} {\overset{\leftrightarrow}{D}}_{\alpha_3} s_b \big] (\bar{q}_c \gamma_{\alpha_2} C \bar{s}_d^T)
\label{def:eta1}
+ (q_a^T C \gamma_{\alpha_1} s_b)  \big[ \bar{q}_c \gamma_{\alpha_2} C {\overset{\leftrightarrow}{D}}_{\alpha_3} \bar{s}_d^T \big]
\Big\} \, ,
\\ \eta^{2}_{\alpha_1\alpha_2\alpha_3} &=&
(\delta^{ac} \delta^{bd} + \delta^{ad} \delta^{bc} ) \times \mathcal{S}\Big\{
\big[q_a^T C \gamma_{\alpha_1} {\overset{\leftrightarrow}{D}}_{\alpha_3} s_b \big] (\bar{q}_c \gamma_{\alpha_2} C \bar{s}_d^T)
\label{def:eta1}
+ (q_a^T C \gamma_{\alpha_1} s_b)  \big[ \bar{q}_c \gamma_{\alpha_2} C {\overset{\leftrightarrow}{D}}_{\alpha_3} \bar{s}_d^T \big]
\Big\} \, ,
\\ \eta^{3}_{\alpha_1\alpha_2\alpha_3} &=&
\epsilon^{abe} \epsilon^{cde} \times \mathcal{S}\Big\{
\big[q_a^T C \gamma_{\alpha_1}\gamma_5 {\overset{\leftrightarrow}{D}}_{\alpha_3} s_b \big] (\bar{q}_c \gamma_{\alpha_2}\gamma_5 C \bar{s}_d^T)
\label{def:eta2}
+ (q_a^T C \gamma_{\alpha_1}\gamma_5 s_b)  \big[ \bar{q}_c \gamma_{\alpha_2}\gamma_5 C {\overset{\leftrightarrow}{D}}_{\alpha_3} \bar{s}_d^T \big]
\Big\} \, ,
\\ \eta^{4}_{\alpha_1\alpha_2\alpha_3} &=&
(\delta^{ac} \delta^{bd} + \delta^{ad} \delta^{bc} ) \times \mathcal{S} \Big\{
\big[q_a^T C \gamma_{\alpha_1}\gamma_5 {\overset{\leftrightarrow}{D}}_{\alpha_3} s_b \big] (\bar{q}_c \gamma_{\alpha_2}\gamma_5 C \bar{s}_d^T)
\label{def:eta2}
+ (q_a^T C \gamma_{\alpha_1}\gamma_5 s_b)  \big[ \bar{q}_c \gamma_{\alpha_2}\gamma_5 C {\overset{\leftrightarrow}{D}}_{\alpha_3} \bar{s}_d^T \big]
\Big\} \, ,
\\ \eta^{5}_{\alpha_1\alpha_2\alpha_3} &=&
\epsilon^{abe} \epsilon^{cde} \times g^{\mu\nu} \mathcal{S}\Big\{
\big[q_a^T C \sigma_{\alpha_1\mu} {\overset{\leftrightarrow}{D}}_{\alpha_3} s_b \big] (\bar{q}_c \sigma_{\alpha_2\nu} C \bar{s}_d^T)
\label{def:eta3}
+ (q_a^T C \sigma_{\alpha_1\mu} s_b)  \big[ \bar{q}_c \sigma_{\alpha_2\nu} C {\overset{\leftrightarrow}{D}}_{\alpha_3} \bar{s}_d^T \big]
\Big\} \, ,
\\ \eta^{6}_{\alpha_1\alpha_2\alpha_3} &=&
(\delta^{ac} \delta^{bd} + \delta^{ad} \delta^{bc} ) \times g^{\mu\nu} \mathcal{S}\Big\{
\big[q_a^T C \sigma_{\alpha_1\mu} {\overset{\leftrightarrow}{D}}_{\alpha_3} s_b \big] (\bar{q}_c \sigma_{\alpha_2\nu} C \bar{s}_d^T)
\label{def:eta3}
+ (q_a^T C \sigma_{\alpha_1\mu} s_b)  \big[ \bar{q}_c \sigma_{\alpha_2\nu} C {\overset{\leftrightarrow}{D}}_{\alpha_3} \bar{s}_d^T \big]
\Big\} \, ,
\end{eqnarray}
\end{widetext}
%
where $\mathcal{S}$ denotes symmetrization and subtracting the trace terms in the set $\{\alpha_1\alpha_2\alpha_3\}$. Three of them $\eta^{1,3,5}_{\alpha_1\alpha_2\alpha_3}$ have the antisymmetric color structure $(qs)_{\mathbf{\bar 3}_C}(\bar q \bar s)_{\mathbf{3}_C}$, and the other three $\eta^{2,4,6}_{\alpha_1\alpha_2\alpha_3}$ have the symmetric color structure $(qs)_{\mathbf{6}_C}(\bar q \bar s)_{\mathbf{\bar 6}_C}$. Considering that the diquark fields $s_a^T C \gamma_\mu s_b/s_a^T C \gamma_\mu \gamma_5 s_b/s_a^T C \sigma_{\mu\nu} s_b$ have the quantum numbers $J^P = 1^+/1^-/1^\pm$ respectively, the first current $\eta^{1}_{\alpha_1\alpha_2\alpha_3}$ has the most stable internal structure and may lead to the best sum rule result.

Besides the above diquark-antidiquark currents, we can construct six color-singlet-color-singlet mesonic-mesonic currents of $J^{PC} = 3^{-+}$:
%
\begin{eqnarray}
\xi^{1}_{\alpha_1\alpha_2\alpha_3} &=&
\label{def:xi1}
\mathcal{S}\Big\{
(\bar{q}_a \gamma_{\alpha_1} q_a) {\overset{\leftrightarrow}{D}}_{\alpha_3}(\bar{s}_b \gamma_{\alpha_2} s_b) \Big\} \, ,
\\ \xi^{2}_{\alpha_1\alpha_2\alpha_3} &=&
\label{def:xi2}
\mathcal{S}\Big\{
(\bar{q}_a \gamma_{\alpha_1} \gamma_5 q_a) {\overset{\leftrightarrow}{D}}_{\alpha_3}(\bar{s}_b \gamma_{\alpha_2} \gamma_5 s_b) \Big\} \, ,
\\ \xi^{3}_{\alpha_1\alpha_2\alpha_3} &=&
g^{\mu\nu} \mathcal{S}\Big\{
(\bar{q}_a \sigma_{\alpha_1\mu} q_a) {\overset{\leftrightarrow}{D}}_{\alpha_3}(\bar{s}_b \sigma_{\alpha_2\nu} s_b)
\label{def:xi3}
\Big\} \, ,
\\ \nonumber \xi^{4}_{\alpha_1\alpha_2\alpha_3} &=&
\mathcal{S}\Big\{
\big[ \bar{q}_a \gamma_{\alpha_1} {\overset{\leftrightarrow}{D}}_{\alpha_3} s_a \big] (\bar{s}_b \gamma_{\alpha_2} q_b)
\label{def:xi4}
\\ \nonumber && ~~~~~~~~~~~ - (\bar{q}_a \gamma_{\alpha_1} s_a) \big [\bar{s}_b \gamma_{\alpha_2} {\overset{\leftrightarrow}{D}}_{\alpha_3}  q_b \big]
\Big\} \, ,
\\ \xi^{5}_{\alpha_1\alpha_2\alpha_3} &=&
\mathcal{S}\Big\{
\big[ \bar{q}_a \gamma_{\alpha_1} \gamma_5 {\overset{\leftrightarrow}{D}}_{\alpha_3} s_a \big] (\bar{s}_b \gamma_{\alpha_2} \gamma_5 q_b)
\label{def:xi5}
\\ \nonumber && ~~~~~~~~~~~ - (\bar{q}_a \gamma_{\alpha_1} \gamma_5 s_a) \big [\bar{s}_b \gamma_{\alpha_2} \gamma_5 {\overset{\leftrightarrow}{D}}_{\alpha_3}  q_b \big]
\Big\} \, ,
\\ \xi^{6}_{\alpha_1\alpha_2\alpha_3} &=&
g^{\mu\nu} \mathcal{S}\Big\{
\big[ \bar{q}_a \sigma_{\alpha_1\mu} {\overset{\leftrightarrow}{D}}_{\alpha_3} s_a \big] (\bar{s}_b \sigma_{\alpha_2\nu} q_b)
\label{def:xi6}
\\ \nonumber && ~~~~~~~~~~~ - (\bar{q}_a \sigma_{\alpha_1\mu} s_a) \big [\bar{s}_b \sigma_{\alpha_2\nu} {\overset{\leftrightarrow}{D}}_{\alpha_3}  q_b \big]
\Big\} \, .
\end{eqnarray}
%
The former three $\xi^{1,2,3}_{\alpha_1\alpha_2\alpha_3}$ have the quark combination $[\bar q q][\bar s s]$, and the derivatives are between the two quark-antiquark pairs; the latter three $\xi^{4,5,6}_{\alpha_1\alpha_2\alpha_3}$ have the quark combination $[\bar q s][\bar s q]$, and the derivatives are inside the quark-antiquark pairs. This difference is useful when investigating their decay properties, which will be discussed in Sec.~\ref{sec:summary}.

We can further construct six color-octet-color-octet mesonic-mesonic currents, which can be related to the above color-singlet-color-singlet mesonic-mesonic currents through the Fierz transformation. Moreover, we can apply the Fierz transformation to derive the relations between diquark-antidiquark and mesonic-mesonic currents:
%
\begin{eqnarray}
\label{eq:fierz}
&& \left(\begin{array}{c}
\eta^{1}_{\alpha_1\alpha_2\alpha_3}
\\
\eta^{2}_{\alpha_1\alpha_2\alpha_3}
\\
\eta^{3}_{\alpha_1\alpha_2\alpha_3}
\\
\eta^{4}_{\alpha_1\alpha_2\alpha_3}
\\
\eta^{5}_{\alpha_1\alpha_2\alpha_3}
\\
\eta^{6}_{\alpha_1\alpha_2\alpha_3}
\end{array}\right)
=
\\ \nonumber &&
\left(\begin{array}{cccccc}
-{1\over2} &  {1\over2} &  {1\over2} & -{1\over2} &  {1\over2} &  {1\over2}
\\
-{1\over2} &  {1\over2} &  {1\over2} &  {1\over2} & -{1\over2} & -{1\over2}
\\
 {1\over2} & -{1\over2} &  {1\over2} & -{1\over2} &  {1\over2} & -{1\over2}
\\
 {1\over2} & -{1\over2} &  {1\over2} &  {1\over2} & -{1\over2} &  {1\over2}
\\
 1         & 1          &  0         &  1         &  1         & 0
\\
 1         & 1          &  0         & -1         & -1         & 0
\end{array}\right)
\left(\begin{array}{c}
\xi^{1}_{\alpha_1\alpha_2\alpha_3}
\\
\xi^{2}_{\alpha_1\alpha_2\alpha_3}
\\
\xi^{3}_{\alpha_1\alpha_2\alpha_3}
\\
\xi^{4}_{\alpha_1\alpha_2\alpha_3}
\\
\xi^{5}_{\alpha_1\alpha_2\alpha_3}
\\
\xi^{6}_{\alpha_1\alpha_2\alpha_3}
\end{array}\right)
 \, .
\end{eqnarray}
%
Therefore, these two constructions are equivalent, and in the following we shall only use $\eta^{1 \cdots 6}_{\alpha_1\alpha_2\alpha_3}$ to perform QCD sum rule analyses. Note that this equivalence is just between diquark-antidiquark and mesonic-mesonic currents, while compact diquark-antidiquark tetraquark states and weakly-bound meson-meson molecular states are totally different. To exactly describe them, one needs non-local interpolating currents, but we are still not capable of using such currents to perform QCD sum rule analyses.

%
\section{QCD sum rule Analysis}
\label{sec:sumrule}
%

In this section we use the currents $\eta^{1 \cdots 6}_{\alpha_1\alpha_2\alpha_3}$ to perform QCD sum rule analyses. We assume that they couple to some exotic state $X$ through
\begin{eqnarray}
\label{eq:defg}
\langle 0| \eta_{\alpha_1\alpha_2\alpha_3} | X \rangle = f_X \epsilon_{\alpha_1\alpha_2\alpha_3} \, ,
\end{eqnarray}
where $f_X$ is the decay constant and $\epsilon_{\alpha_1\alpha_2\alpha_3}$ is the traceless and symmetric polarization tensor, satisfying:
\begin{eqnarray}
\epsilon_{\alpha_1\alpha_2\alpha_3} \epsilon^*_{\beta_1\beta_2\beta_3} &=& \mathcal{S}^\prime [\tilde g_{\alpha_1 \beta_1} \tilde g_{\alpha_2 \beta_2} \tilde g_{\alpha_3 \beta_3}] \, .
\end{eqnarray}
In this expression $\tilde g_{\mu \nu} = g_{\mu \nu} - q_\mu q_\nu / q^2$, and $\mathcal{S}^\prime$ denotes symmetrization and subtracting the trace
terms in the sets $\{\alpha_1\alpha_2\alpha_3\}$ and $\{\beta_1\beta_2\beta_3\}$.

Based on Eq.~(\ref{eq:defg}), we study the two-point correlation function
%
\begin{eqnarray}
&& \Pi_{\alpha_1\alpha_2\alpha_3,\beta_1\beta_2\beta_3}(q^2)
\label{def:pi}
\\ \nonumber &\equiv& i \int d^4x e^{iqx} \langle 0 | {\bf T}[ \eta_{\alpha_1\alpha_2\alpha_3}(x) \eta^{\dagger}_{\beta_1\beta_2\beta_3} (0)] | 0 \rangle
\\ \nonumber &=& (-1)^J~\mathcal{S}^\prime [\tilde g_{\alpha_1 \beta_1} \tilde g_{\alpha_2 \beta_2} \tilde g_{\alpha_3 \beta_3}]~\Pi (q^2) \, ,
\end{eqnarray}
%
at both hadron and quark-gluon levels.

At the hadron level we use the dispersion relation to express Eq.~(\ref{def:pi}) as:
%
\begin{equation}
\Pi(q^2) = \int^\infty_{4 m_s^2}\frac{\rho(s)}{s-q^2-i\varepsilon}ds \, ,
\label{eq:hadron}
\end{equation}
%
where $\rho(s)$ is the spectral density. Then we parameterize it using one
pole dominance for the ground state $X$ and a continuum contribution:
%
\begin{eqnarray}
\rho(s) &\equiv& \sum_n\delta(s-M^2_n) \langle 0| \eta | n\rangle \langle n| {\eta^{\dagger}} |0 \rangle
\\ \nonumber &=& f^2_X \delta(s-M^2_X) + \rm{continuum} \, .
\label{eq:rho}
\end{eqnarray}
%

At the quark-gluon level we insert $\eta^{1 \cdots 6}_{\alpha_1\alpha_2\alpha_3}$ into Eq.~(\ref{def:pi}) and calculate it using the method of operator product expansion (OPE). After performing the Borel transformation to Eq.~(\ref{def:pi}) at both hadron and quark-gluon levels, we can approximate the continuum using the spectral density above a threshold value $s_0$, and obtain the sum rule equation
%
\begin{equation}
\Pi(s_0, M_B^2) \equiv f^2_X e^{-M_X^2/M_B^2} = \int^{s_0}_{4 m_s^2} e^{-s/M_B^2}\rho(s)ds \, .
\label{eq_fin}
\end{equation}
%
We can use it to further evaluate $M_X$, the mass of $X$, through,
%
\begin{eqnarray}
M^2_X(s_0, M_B) &=& \frac{\frac{\partial}{\partial(-1/M_B^2)}\Pi(s_0, M_B^2)}{\Pi(s_0, M_B^2)}
\label{eq:LSR}
\\ \nonumber &=& \frac{\int^{s_0}_{4 m_s^2} e^{-s/M_B^2}s\rho(s)ds}{\int^{s_0}_{4 m_s^2} e^{-s/M_B^2}\rho(s)ds} \, .
\end{eqnarray}
%

%
\begin{widetext}
In the present study we have calculated OPEs up to the tenth dimension, including the perturbative term, the strange quark mass, the gluon condensate, the quark condensate, the quark-gluon mixed condensate, and their combinations:
\begin{eqnarray}
 \label{eq:pieta1}
 \Pi_{11} &=& \int^{s_0}_{4 m_s^2} \Bigg [
{s^5 \over 691200 \pi^6}-{m_s^2 s^4 \over 14336 \pi^6}
+ \Big(- {179 \langle g_s^2 GG \rangle \over 5806080 \pi^6}+ {m_s^4 \over 2016 \pi^6}-{m_s \langle \bar q q \rangle \over 720 \pi^4} +{m_s \langle \bar s s \rangle
\over 1512 \pi^4}\Big ) s^3\\ \nonumber &&
+ \Big ( {37 \langle g_s^2 GG \rangle m_s^2 \over 122880 \pi^6 }
- {91 m_s \langle g_s \bar q \sigma G q \rangle \over 30720 \pi^4}+ {m_s^3 \langle \bar q q \rangle \over80\pi^4}-{m_s^3\langle \bar s s \rangle \over240 \pi^4}+{\langle \bar q q\rangle \langle \bar s s\rangle \over 60\pi^2}\Big )s^2
\\ \nonumber &&
+ \Big(-{\langle g_s^2 GG \rangle m_s^4 \over18432 \pi^6 }
+{3m_s^3 \langle g_s \bar q \sigma G q \rangle \over256\pi^4 }
+{5\langle g_s^2 GG \rangle m_s \langle \bar q q \rangle \over3456\pi^4}
-{7\langle g_s^2 GG \rangle m_s \langle \bar s s \rangle \over 8640 \pi^4}
+{5 \langle g_s \bar q \sigma G q \rangle \langle \bar s s \rangle \over 288 \pi^2}
-{m_s^2\langle \bar q q\rangle \langle \bar s s\rangle \over 12 \pi^2}
\\ \nonumber &&
+{5\langle \bar q q \rangle \langle g_s \bar s \sigma G s \rangle \over 288 \pi^2}\Big)s
+\Big ( {\langle g_s^2 GG \rangle m_s \langle g_s \bar q \sigma G q \rangle \over 4608 \pi^4 }
- { m_s^2 \langle g_s\bar q \sigma G q  \rangle \langle \bar q q \rangle \over 12 \pi^2}
+ { \langle g_s^2 GG \rangle  m_s^3 \langle \bar s s\rangle  \over 13824 \pi^4 }
- { 3 m_s^2\langle g_s \bar q \sigma G q \rangle \langle \bar s s \rangle \over 128\pi^2 }
\\ \nonumber &&
-{\langle g_s^2 GG \rangle \langle \bar q q \rangle \langle \bar s s \rangle \over 324 \pi^2 }
+ { 17\langle g_s \bar q \sigma G q \rangle \langle g_s \bar s \sigma G s \rangle\over 3456 \pi^2} -{m_s^2 \langle \bar q q\rangle \langle g_s \bar s \sigma G s \rangle \over 576 \pi^2}\Big ) \Bigg ] e^{-s/M^2} ds
\\ \nonumber &&
+ \Big ( -{m_s^2 {\langle g_s\bar q \sigma G q  \rangle}^2   \over 24 \pi^2}
+{ 2m_s \langle g_s \bar q \sigma G q\rangle \langle \bar qq \rangle \langle \bar s s \rangle \over 9  } \Big )
\, ,
\\ \label{eq:pieta2} \Pi_{22} &=& \int^{s_0}_{4 m_s^2} \Bigg [ {s^5 \over 345600 \pi^6}
-  {m_s^2 \over 7168 \pi^6}s^4
+  \Big (- {199 \langle g_s^2 GG \rangle \over 5806080 \pi^6 }
+ {m_s^4 \over 1008 \pi^6}
-  {m_s \langle \bar q q \rangle \over 360 \pi^4}
+ {m_s \langle \bar s s \rangle \over 756 \pi^4}\Big ) s^3
\\ \nonumber &&
+ \Big ( { 41 m_s^2 \langle g_s^2 GG \rangle \over 122880 \pi^6 }
-  { 239 m_s \langle g_s \bar q \sigma G q \rangle \over 30720 \pi^4 }
+ {m_s^3  \langle \bar q q \rangle \over 40 \pi^4}
-  {m_s^3  \langle \bar s s  \rangle \over 120 \pi^4}
+ { \langle \bar q q \rangle \langle \bar s s  \rangle \over 30 \pi^2}\Big )s^2
\\ \nonumber &&
+ \Big(- { 5 m_s^4 \langle g_s^2 GG \rangle \over 18432 \pi^6 }
+ { 7 m_s^3 \langle g_s \bar q \sigma G q \rangle \over 256 \pi^4 }
-  { 5 m_s \langle \bar q q \rangle \langle g_s^2 GG \rangle \over 3456 \pi^2}
-  { m_s \langle \bar s s \rangle \langle g_s^2 GG \rangle \over 1080 \pi^4}
+ { 13 \langle \bar s s \rangle \langle g_s \bar q \sigma G q \rangle \over 288 \pi^2}
-  { m_s^2 \langle \bar s s \rangle \langle \bar q q \rangle \over 6 \pi^2}
\\ \nonumber &&
+ { 13 \langle \bar q q \rangle \langle g_s \bar s \sigma G s \rangle \over 288 \pi^2 } \Big ) s
+\Big ( { 49 \langle g_s \bar q \sigma G q \rangle \langle g_s \bar s \sigma G s \rangle \over 3456 \pi^2 }
- { 5 m_s^2  \langle \bar q q \rangle \langle g_s \bar s \sigma G s \rangle \over 576 \pi^2}
-  { m_s \langle g_s^2 GG \rangle \langle g_s \bar q \sigma G q \rangle \over 4608 \pi^4 }
-  { m_s^2 \langle \bar q q \rangle \langle g_s \bar q \sigma G q \rangle \over 6 \pi^2 }
\\ \nonumber &&
+ { 5 m_s^3 \langle \bar s s \rangle \langle g_s^2 GG \rangle \over 13824 \pi^4 }
-  { 7 m_s^2 \langle \bar s s \rangle \langle g_s \bar q \sigma G q \rangle \over 128 \pi^2 }
+ { \langle \bar q q \rangle \langle \bar s s \rangle \langle g_s^2 GG \rangle \over 324 \pi^2 }\Big ) \Bigg ] e^{-s/M_B^2} ds
\\ \nonumber &&
+ \Big ( - { m_s^2 \langle g_s \bar q \sigma G q \rangle^2 \over 12 \pi^2}
+ { 4 m_s \langle \bar s s \rangle \langle \bar q q \rangle \langle g_s \bar q \sigma G q \rangle \over 9 } \Big )
\, ,
\\  \label{eq:pieta3}
\Pi_{33} &=& \int^{s_0}_{4 m_s^2} \Bigg [+
{s^5 \over 691200 \pi^6}-{m_s^2 s^4 \over 14336 \pi^6}
+ \Big(- {179 \langle g_s^2 GG \rangle \over 5806080 \pi^6}+ {m_s^4 \over 2016 \pi^6}+{m_s \langle \bar q q \rangle \over 720 \pi^4} +{m_s \langle \bar s s \rangle \over 1512 \pi^4}\Big ) s^3
\\ \nonumber &&
+ \Big ( {37 \langle g_s^2 GG \rangle m_s^2 \over 122880 \pi^6 }
 +{91 m_s \langle g_s \bar q \sigma G q \rangle \over 30720 \pi^4}
 -{m_s^3 \langle \bar q q \rangle \over80\pi^4}
 -{m_s^3\langle \bar s s \rangle \over240 \pi^4}
 -{\langle \bar q q\rangle \langle \bar s s\rangle \over 60\pi^2}\Big )s^2
\\ \nonumber &&
+ \Big(-{\langle g_s^2 GG \rangle m_s^4 \over18432 \pi^6 }
-{3m_s^3 \langle g_s \bar q \sigma G q \rangle \over256\pi^4 }
-{5\langle g_s^2 GG \rangle m_s \langle \bar q q \rangle \over3456\pi^4}
-{7\langle g_s^2 GG \rangle m_s \langle \bar s s \rangle \over 8640 \pi^4}
-{5 \langle g_s \bar q \sigma G q \rangle \langle \bar s s \rangle \over 288 \pi^2}
+{m_s^2\langle \bar q q\rangle \langle \bar s s\rangle \over 12 \pi^2}
\\ \nonumber &&
-{5\langle \bar q q \rangle \langle g_s \bar s \sigma G s \rangle \over 288 \pi^2}\Big)s
+\Big ( -{\langle g_s^2 GG \rangle m_s \langle g_s \bar q \sigma G q \rangle \over 4608 \pi^4 }
- { m_s^2 \langle g_s\bar q \sigma G q  \rangle \langle \bar q q \rangle \over 12 \pi^2}
+ { \langle g_s^2 GG \rangle  m_s^3 \langle \bar s s\rangle  \over 13824 \pi^4 }
 +{ 3 m_s^2\langle g_s \bar q \sigma G q \rangle \langle \bar s s \rangle \over 128\pi^2 }
\\ \nonumber &&
+ {\langle g_s^2 GG \rangle \langle \bar q q \rangle \langle \bar s s \rangle \over 324 \pi^2 }
-{ 17\langle g_s \bar q \sigma G q \rangle \langle g_s \bar s \sigma G s \rangle\over 3456 \pi^2} +{m_s^2 \langle \bar q q\rangle \langle g_s \bar s \sigma G s \rangle \over 576 \pi^2}\Big ) \Bigg ] e^{-s/M^2} ds
\\ \nonumber &&
+ \Big ( -{m_s^2 {\langle g_s\bar q \sigma G q  \rangle}^2   \over 24 \pi^2}
+{ 2m_s \langle g_s \bar q \sigma G q\rangle \langle \bar qq \rangle \langle \bar s s \rangle \over 9  } \Big )
\, ,
\\ \label{eq:pieta4}
\Pi_{44} &=& \int^{s_0}_{4 m_s^2} \Bigg [ {s^5 \over 345600 \pi^6}
-  { m_s^2 s^4 \over 7168 \pi^6 }
+ \Big (- { 199 \langle g_s^2 GG \rangle \over 5806080 \pi^6 }
+ { m_s^4 \over 1008 \pi^6}
+ { m_s \langle \bar q q \rangle \over 360 \pi^4 }
+ { m_s \langle \bar s s \rangle \over 756 \pi^4 } \Big ) s^3
\\ \nonumber &&
+ \Big ( { 41 m_s^2 \langle g_s^2 GG \rangle \over 122880 \pi^6 }
+ { 239 m_s \langle g_s \bar q \sigma G q \rangle \over 30720 \pi^4}
-  { m_s^3 \langle \bar q q \rangle \over 40 \pi^4 }
-  { m_s^3 \langle \bar s s \rangle \over 120 \pi^4 }
-  { \langle \bar q q \rangle \langle \bar s s \rangle \over 30 \pi^2 } \Big ) s^2
\\ \nonumber &&
+ \Big ( - { 5 m_s^4 \langle g_s^2 GG \rangle \over 18432 \pi^6 }
-  { 7 m_s^3  \langle g_s \bar q \sigma G q \rangle \over 256 \pi^4 }
+ { 5 m_s \langle \bar q q \rangle \langle g_s^2 GG \rangle \over 3456 \pi^4 }
-  { m_s \langle \bar s s \rangle \langle g_s^2 GG \rangle \over 1080 \pi^4 }
-  { 13 \langle \bar s s \rangle \langle g_s \bar q \sigma G q \rangle \over 288 \pi^2 }
+ { m_s^2 \langle \bar s s \rangle \langle \bar q q \rangle \over 6 \pi^2 }
\\ \nonumber &&
-  { 13 \langle \bar q q \rangle \langle g_s \bar s \sigma G s \rangle \over 288 \pi^2 } \Big ) s
+ \Big ( { m_s \langle g_s^2 GG \rangle \langle g_s \bar q \sigma G q \rangle \over 4608 \pi^4}
+ { 5 m_s^3 \langle g_s^2 GG \rangle \langle \bar s s \rangle \over 13824 \pi^4}
- { m_s^2 \langle \bar q q \rangle \langle g_s \bar q \sigma G q \rangle \over 6 \pi^2 }
+ { 7 m_s^2 \langle \bar s s \rangle \langle g_s \bar q \sigma G q \rangle \over 128 \pi^2}
\\ \nonumber &&
-  { \langle \bar s s \rangle \langle \bar q q \rangle \langle g_s^2 GG \rangle \over 324 \pi^2 }
-  { 49 \langle g_s \bar q \sigma G q \rangle \langle g_s \bar s \sigma G s \rangle \over 3456 \pi^2 }
+ { 5 m_s^2 \langle \bar q q \rangle \langle g_s \bar s \sigma G s \rangle \over 576 \pi^2 } \Big ) \Bigg ] e^{-s/M_B^2} ds
\\  \nonumber &&
+ \Big ( { 4 m_s \langle \bar q q \rangle \langle \bar s s \rangle \langle g_s \bar q \sigma G q \rangle \over 9 }
- { m_s^2 \langle g_s \bar q \sigma G q \rangle^2 \over 12 \pi^2 } \Big )
\, ,
\\ \label{eq:pieta5}
\Pi_{55} &=& \int^{s_0}_{4 m_s^2} \Bigg [
+{s^5\over 345600 \pi^6}-{m_s^2 s^4\over 7168\pi^6}
 +\Big (-{73\langle g_s^2 GG \rangle \over1451520 \pi^6}+{m_s^4\over 1008 \pi^6}
 +{m_s \langle \bar s s \rangle\over 756 \pi^4}  \Big )s^3
 +\Big ({\langle g_s^2 G G \rangle m_s^2 \over 2048\pi^6} -{m_s^3 \langle \bar s s \rangle \over 120 \pi^4}\Big )s^2
 \\ \nonumber &&
 +\Big (-{\langle g_s^2 GG \rangle m_s^4\over 11520 \pi^6}
 -{23 \langle g_s^2 G G\rangle m_s \langle \bar s s \rangle \over 17280\pi^4 }\Big )s
 -{m_s^2\langle g_s \bar q \sigma G q \rangle \langle\bar  s s \rangle \over 6 \pi^2 }
  +{\langle g_s^2 G G \rangle m_s^3 \langle \bar s s \rangle \over 6912\pi^4}\Bigg ] e^{-s/M_B^2} ds
 \\ \nonumber &&
+ \Big ( -{m_s^2 {\langle g_s \bar q \sigma G q \rangle}^2 \over 12 \pi^2 }
+{4m_s \langle g_s \bar q \sigma G q \rangle \langle \bar q q \rangle \langle \bar s s \rangle \over 9} \Big )
\, ,
\\ \label{eq:pieta6}
\Pi_{66} &=& \int^{s_0}_{4 m_s^2} \Bigg [ {s^5 \over 172800 \pi^6}
-  { m_s^2 s^4 \over 3584 \pi^6 }
+ \Big ( - { 25 \langle g_s^2 GG \rangle \over 290304 \pi^6 }
+ { m_s^4 \over 504 \pi^6}
+ { m_s \langle \bar s s \rangle \over 378 \pi^4 } \Big ) s^3
+ \Big ( { 9 m_s^2 \langle g_s^2 GG \rangle \over 10240 \pi^6 }
- { m_s^3 \langle \bar s s \rangle \over 60 \pi^4 } \Big ) s^2
\\ \nonumber &&
+ \Big ( - { m_s^4 \langle g_s^2 GG \rangle \over 2304 \pi^6 }
-  { 43 m_s \langle g_s^2 GG \rangle \langle \bar s s \rangle \over 17280 \pi^4 } \Big ) s
-  { m_s^2 \langle g_s \bar q \sigma G q \rangle \langle \bar q q \rangle \over 3 \pi^2}
+ { 5 m_s^3 \langle g_s^2 GG \rangle \langle \bar s s \rangle \over 6912 \pi^4 } \Bigg ] e^{-s/M_B^2} ds
\\ \nonumber &&
+ \Big ( - { m_s^2 \langle g_s \bar q \sigma G q \rangle^2 \over 6 \pi^2 }
+ { 8 m_s \langle g_s \bar q \sigma G q \rangle \langle \bar q q \rangle \langle \bar s s \rangle \over 9 } \Big )
\, .
\end{eqnarray}
\end{widetext}
Based on these expressions, we shall perform numerical analyses in the next section.

\section{Numerical Analyses}\label{sec:numerical}
%
In this section we use the sum rules given in Eqs.~(\ref{eq:pieta1}-\ref{eq:pieta6}) to perform numerical analyses. The following values are used for various quark and gluon parameters~\cite{Yang:1993bp,Narison:2002pw,Gimenez:2005nt,Jamin:2002ev,Ioffe:2002be,Ovchinnikov:1988gk,Ellis:1996xc,pdg}:
%
\begin{eqnarray}
\nonumber m_s(2\mbox{ GeV}) &=& 96 ^{+8}_{-4} \mbox{ MeV} \, ,
\\ \langle g_s^2GG\rangle &=& (0.48\pm 0.14) \mbox{ GeV}^4 \, ,
\\ \nonumber  \langle\bar qq \rangle &=& -(0.240 \pm 0.010)^3 \mbox{ GeV}^3 \, ,
\\ \nonumber  \langle\bar ss \rangle &=& (0.8\pm 0.1)\times \langle\bar qq \rangle \, ,
\label{condensates}
\\
\nonumber \langle g_s\bar q\sigma G q\rangle &=& - M_0^2\times\langle\bar qq\rangle \, ,
\\
\nonumber \langle g_s\bar s\sigma G s\rangle &=& - M_0^2\times\langle\bar ss\rangle \, ,
\\
\nonumber M_0^2 &=& (0.8 \pm 0.2) \mbox{ GeV}^2 \, .
\end{eqnarray}
%

%
\begin{figure*}[hbt]
\begin{center}
\includegraphics[width=0.32\textwidth]{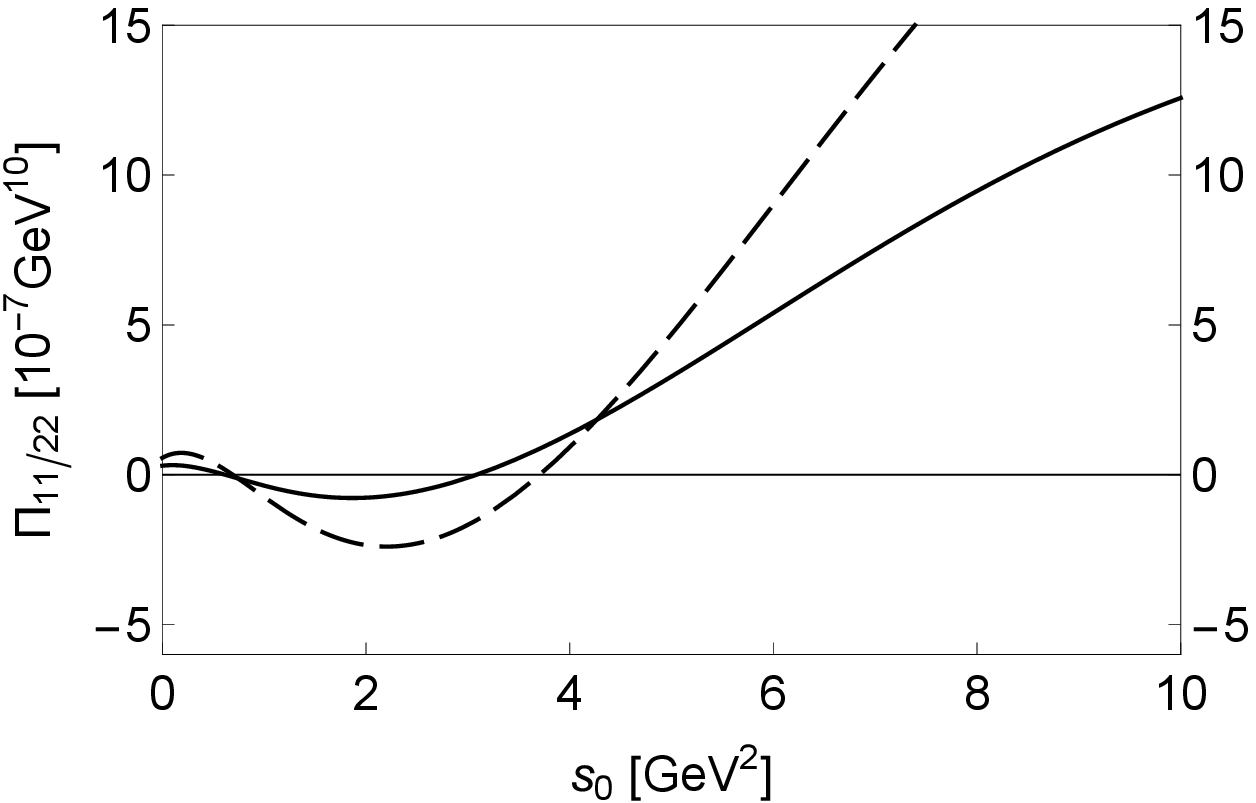}
\includegraphics[width=0.32\textwidth]{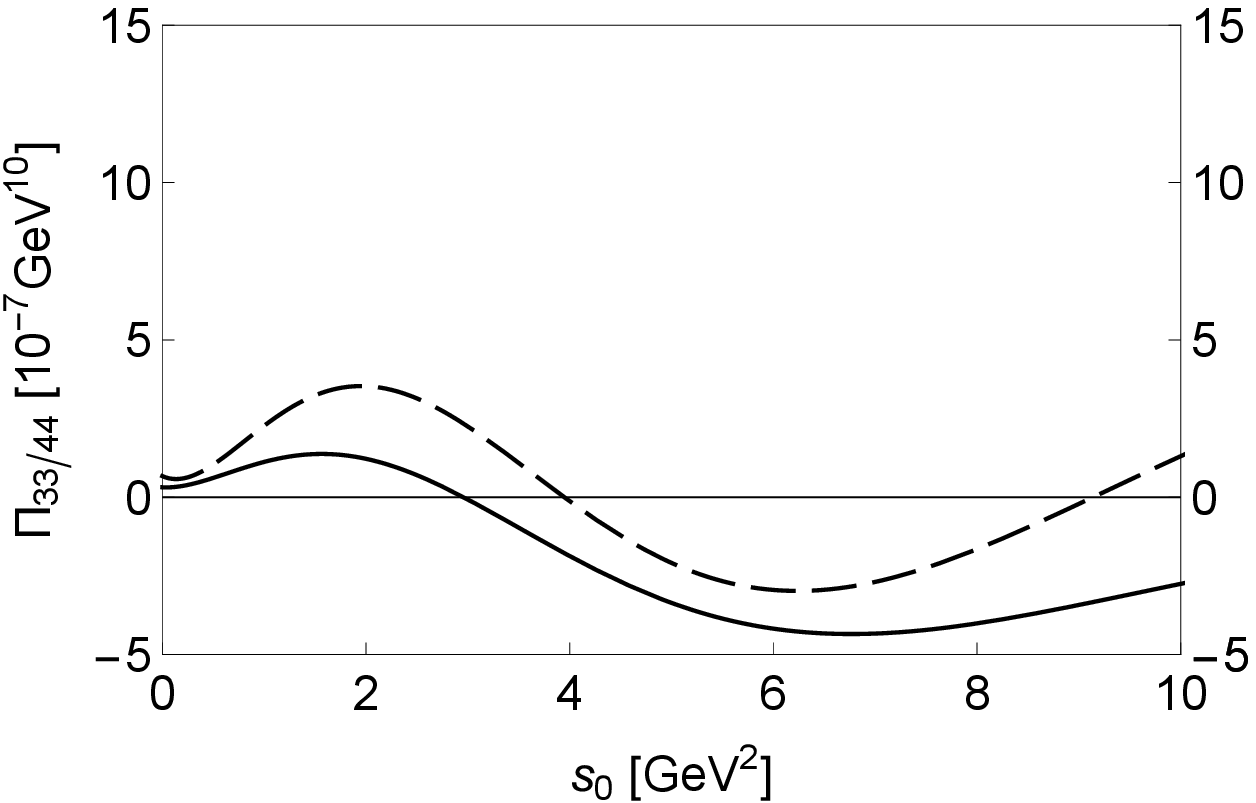}
\includegraphics[width=0.32\textwidth]{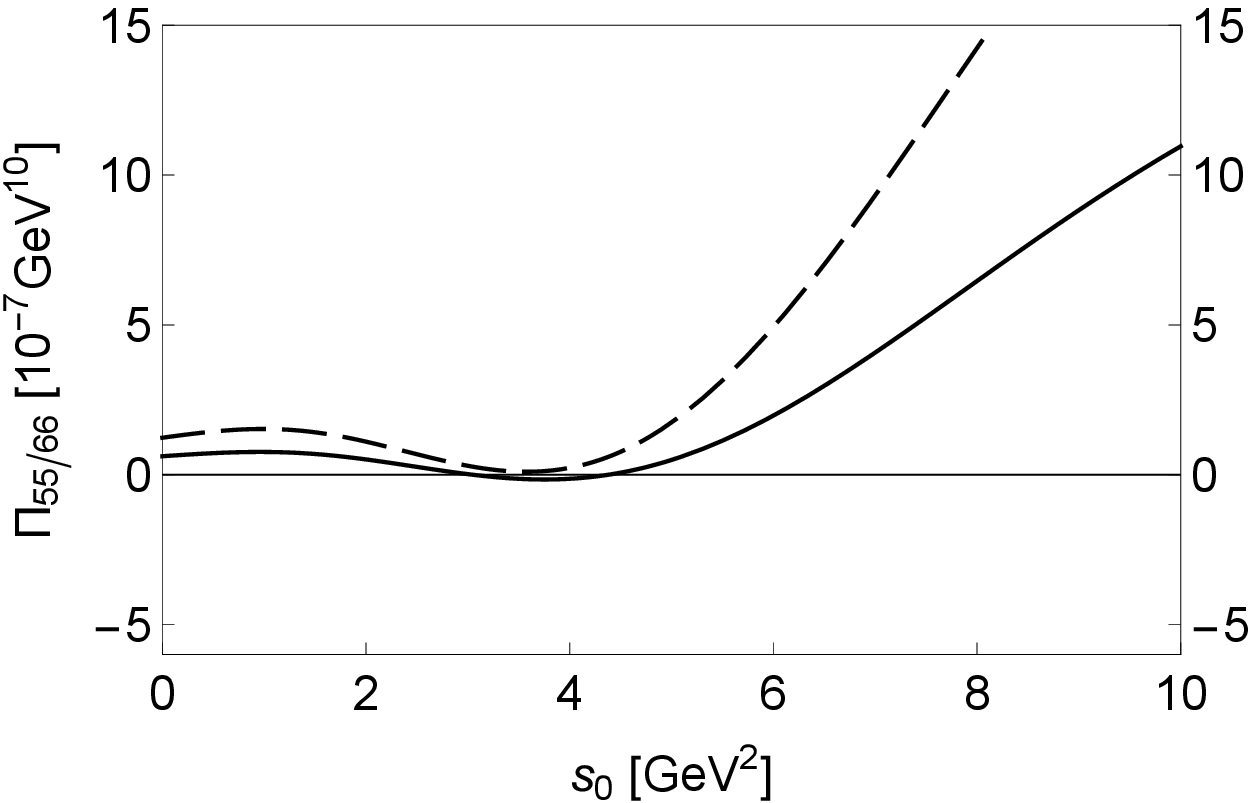}
\caption{The two-point correlation functions, $\Pi_{11}(s_0, M_B^2)$ (left-solid), $\Pi_{22}(s_0, M_B^2)$ (left-dashed), $\Pi_{33}(s_0, M_B^2)$ (middle-solid), $\Pi_{44}(s_0, M_B^2)$ (middle-dashed), $\Pi_{55}(s_0, M_B^2)$ (right-solid), and $\Pi_{66}(s_0, M_B^2)$ (right-dashed), as functions of the threshold value $s_0$. These curves are obtained by setting $M_B^2 = 1.4$~GeV$^2$.}
\label{fig:pi12}
\end{center}
\end{figure*}

To begin with, we show Eqs.~(\ref{eq:pieta1}-\ref{eq:pieta6}) in Fig.~\ref{fig:pi12} as functions of the threshold value $s_0$. We find that $\Pi_{33}(M_B^2, s_0)$ and $\Pi_{44}(M_B^2, s_0)$ are both negative when $s_0$ is around $6$~GeV$^2$. This suggests that they are both non-physical in this energy region, so we shall not investigate $\eta^3_{\alpha_1\alpha_2\alpha_3}$ and $\eta^4_{\alpha_1\alpha_2\alpha_3}$ any more.

The mass $M_X$ depends on two free parameters, the threshold value $s_0$ and the Borel mass $M_B$. To find their proper working regions, we investigate three aspects: a) the OPE convergence, b) the pole contribution, and c) the mass dependence on $M_B$ and $s_0$.

%
\begin{figure}[hbt]
\begin{center}
\includegraphics[width=0.47\textwidth]{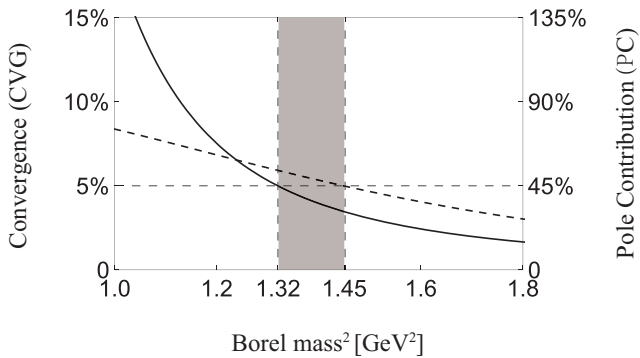}
\caption{CVG (solid curve, defined in Eq.~(\ref{eq:convergence})) and PC (dashed curve, defined in Eq.~(\ref{eq:pole})) as functions of the Borel mass $M_B$. These curves are obtained using the current $\eta^1_{\alpha_1\alpha_2\alpha_3}$ when setting $s_0 = 7.2$~GeV$^2$.}
\label{fig:cvgpole}
\end{center}
\end{figure}

Taking the current $\eta^1_{\alpha_1\alpha_2\alpha_3}$ as an example, whose sum rules are given in Eq.~(\ref{eq:pieta1}). First we investigate the convergence of OPE, which is the cornerstone of a reliable QCD sum rule analysis. We require the $D=10$ term to be less than 5\%:
\begin{eqnarray}
\mbox{CVG} &\equiv& \left|\frac{ \Pi_{11}^{D=10}(s_0, M_B^2) }{ \Pi_{11}(s_0, M_B^2) }\right| \leq 5\% \, .
\label{eq:convergence}
\end{eqnarray}
As shown in Fig.~\ref{fig:cvgpole} using the solid curve, the lower bound of the Borel mass is determined to be $M_B^2 > 1.32$~GeV$^2$, when setting $s_0 = 7.2$~GeV$^2$.

Then we investigate the one-pole-dominance assumption by requiring the pole contribution (PC) to be larger than 45\%:
\begin{eqnarray}
\mbox{PC} &\equiv& \left|\frac{ \Pi_{11}(s_0, M_B^2) }{ \Pi_{11}(\infty, M_B^2) }\right| \geq 45\% \, .
\label{eq:pole}
\end{eqnarray}
As shown in Fig.~\ref{fig:cvgpole} using the dashed curve, the upper bound of the Borel mass is determined to be $M_B^2 < 1.45$~GeV$^2$, when setting $s_0 = 7.2$~GeV$^2$.

Altogether we obtain the Borel window to be $1.32$~GeV$^2 < M_B^2 < 1.45$~GeV$^2$ when setting $s_0 = 7.2$~GeV$^2$. Redoing the same procedures by changing $s_0$, we find that there are non-vanishing Borel windows as long as $s_0 > 6.7$~GeV$^2$.

\begin{figure*}[]
\begin{center}
\includegraphics[width=0.45\textwidth]{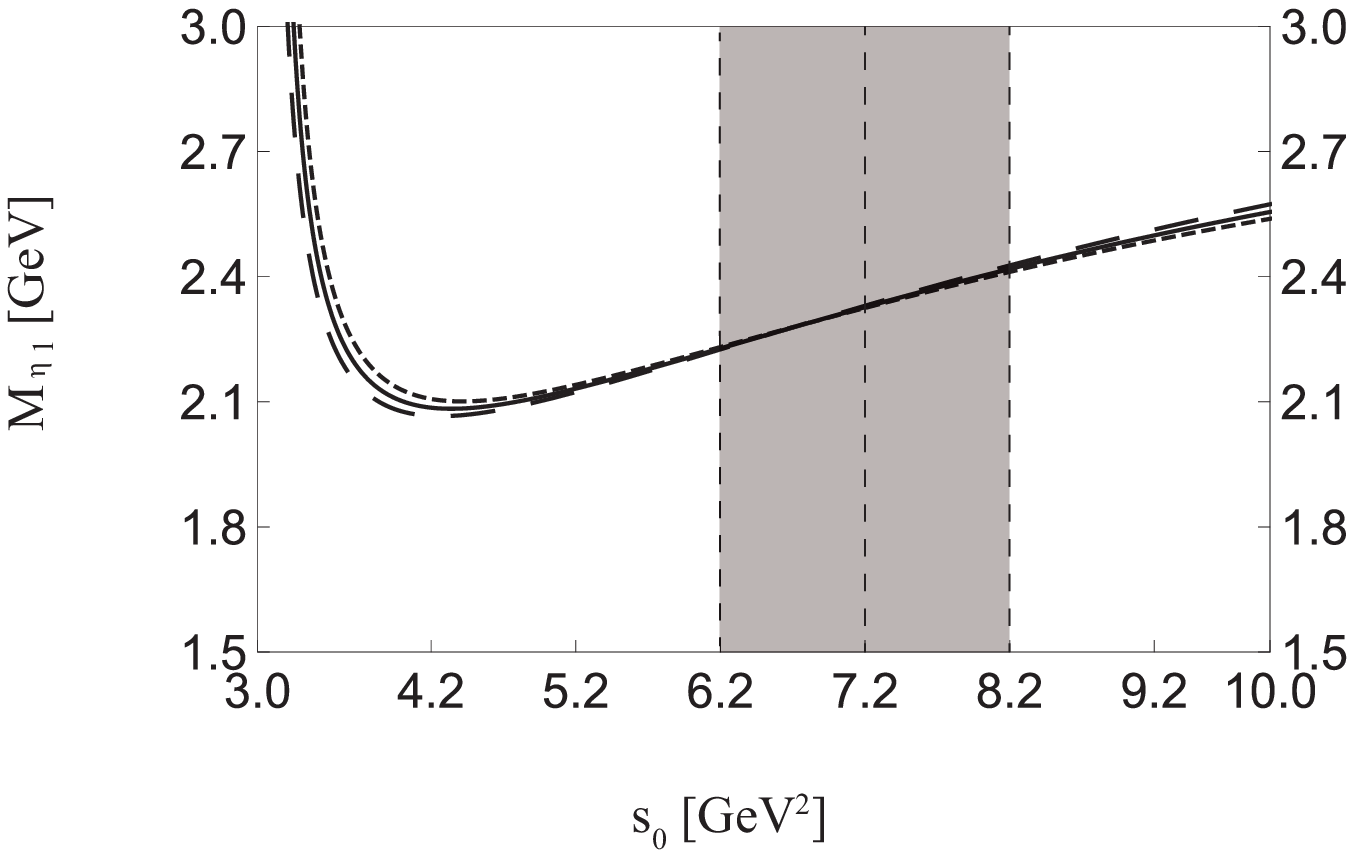}
~~~~~
\includegraphics[width=0.45\textwidth]{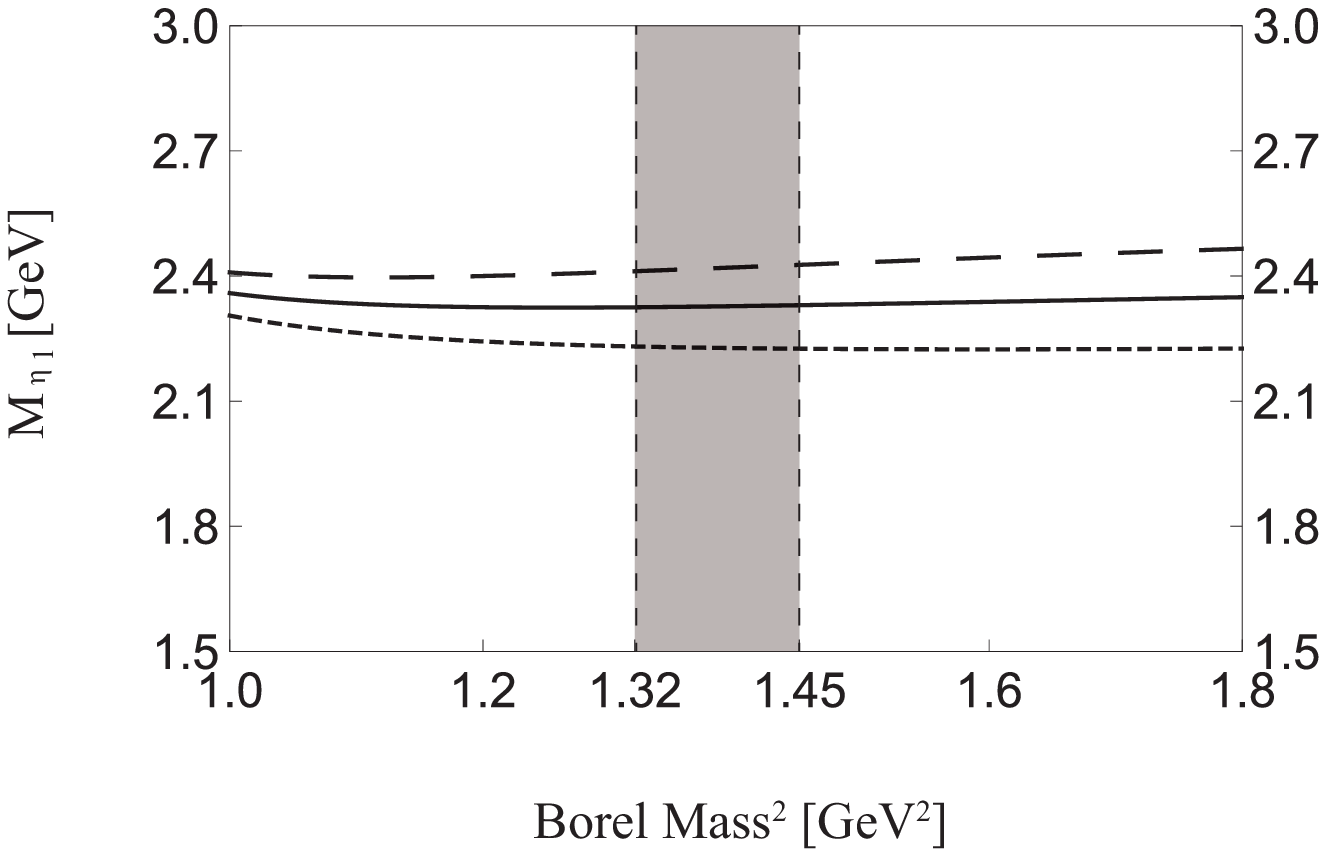}
\caption{
Mass calculated using the current $\eta^{1}_{\alpha_1\alpha_2\alpha_3}$, as a function of the threshold value $s_0$ (left) and the Borel mass $M_B$ (right).
In the left panel the short-dashed/solid/long-dashed curves are depicted when setting $M_B^2 = 1.32/1.38/1.45$ GeV$^2$, respectively.
In the right panel the short-dashed/solid/long-dashed curves are depicted when setting $s_0 = 6.2/7.2/8.2$ GeV$^2$, respectively.}
\label{fig:eta1mass}
\end{center}
\end{figure*}

Finally, we study the mass dependence on $M_B$ and $s_0$. We show the mass $M_X$ in Fig.~\ref{fig:eta1mass} with respective to these two parameters. It is stable around $s_0 \sim 7.2$~GeV$^2$, and its dependence on $M_B$ is weak in the Borel window $1.32$~GeV$^2 < M_B^2 < 1.45$~GeV$^2$. Accordingly, we choose the working regions to be $6.2$~GeV$^2< s_0 < 8.2$~GeV$^2$ and $1.32$~GeV$^2 < M_B^2 < 1.45$~GeV$^2$, where the mass $M_X$ is evaluated to be
\begin{equation}
M_{\eta_1} = 2.33^{+0.19}_{-0.16}{\rm~GeV} \, .
\end{equation}
Here the central value corresponds to $s_0=7.2$~GeV$^2$ and $M_B^2 = 1.38$~GeV$^2$, and the uncertainty comes from $M_B$, $s_0$, and various quark and gluon parameters listed in Eqs.~(\ref{condensates}).

\begin{figure*}[]
\begin{center}
\includegraphics[width=0.45\textwidth]{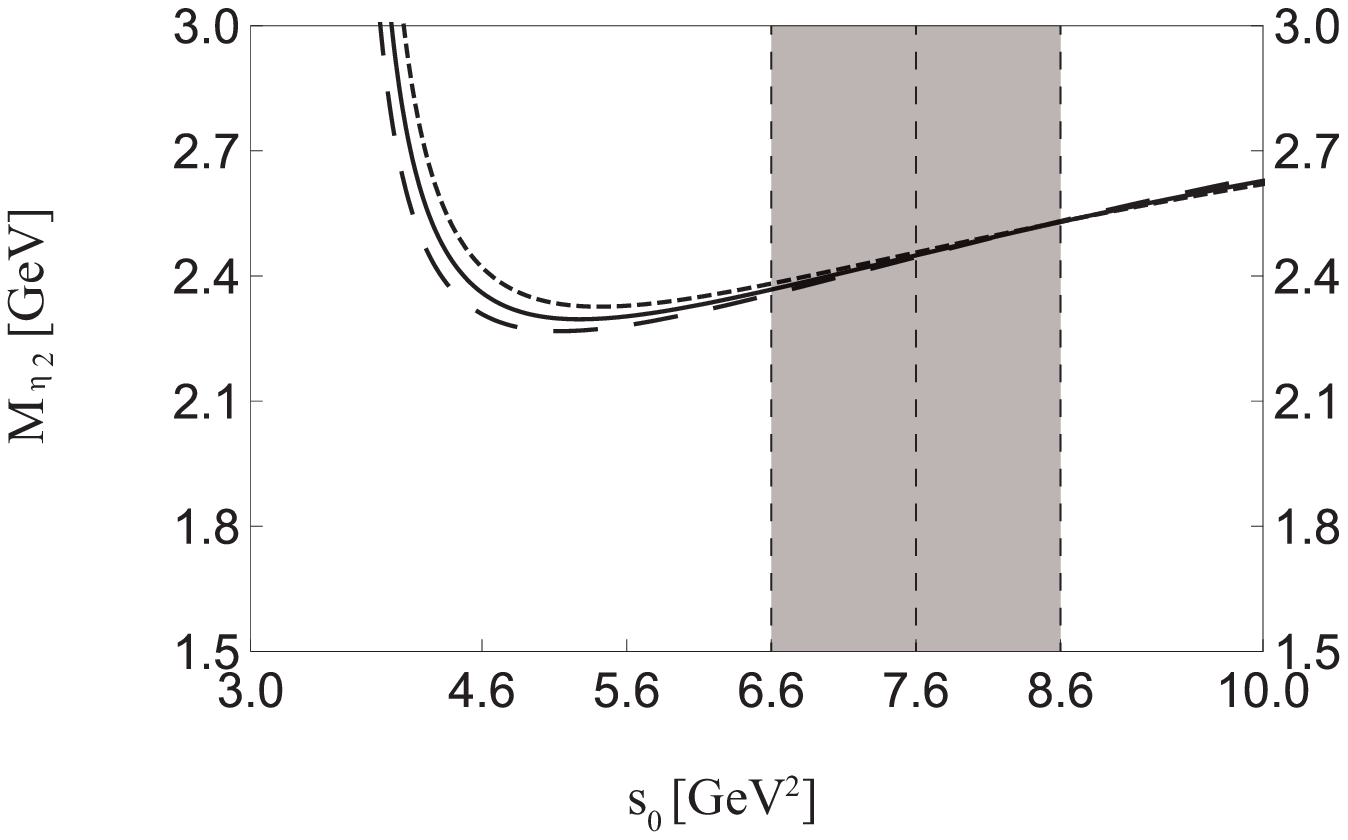}
~~~~~
\includegraphics[width=0.45\textwidth]{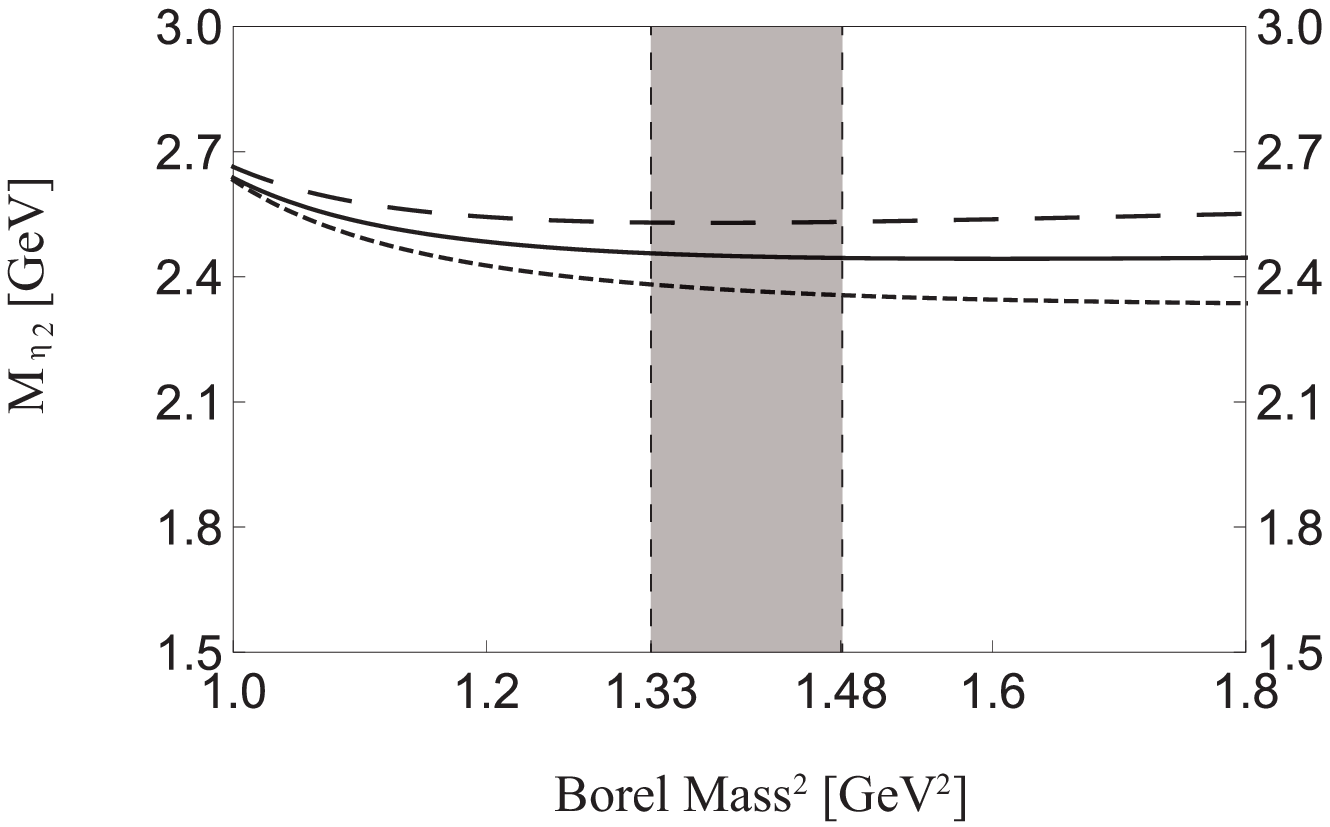}
\caption{
Mass calculated using the current $\eta^{2}_{\alpha_1\alpha_2\alpha_3}$, as a function of the threshold value $s_0$ (left) and the Borel mass $M_B$ (right).
In the left panel the short-dashed/solid/long-dashed curves are depicted when setting $M_B^2 = 1.33/1.40/1.48$ GeV$^2$, respectively.
In the right panel the short-dashed/solid/long-dashed curves are depicted when setting $s_0 = 6.6/7.6/8.6$ GeV$^2$, respectively.}
\label{fig:eta2mass}
\end{center}
\end{figure*}

Similarly, we use $\eta^2_{\alpha_1\alpha_2\alpha_3}$ to perform numerical analyses. We show the mass extracted in Fig.~\ref{fig:eta2mass} as a function of the threshold value $s_0$ (left) and the Borel mass $M_B$ (right). After extracting the working regions to be $6.6$~GeV$^2< s_0 < 8.6$~GeV$^2$ and $1.33$~GeV$^2 < M_B^2 < 1.48$~GeV$^2$, we obtain
\begin{equation}
M_{\eta_2} = 2.45^{+0.27}_{-0.18}{\rm~GeV} \, ,
\end{equation}
where the central value corresponds to $s_0=7.6$~GeV$^2$ and $M_B^2 = 1.40$~GeV$^2$.

The same procedures are applied to analyses the currents $\eta^{5}_{\alpha_1\alpha_2\alpha_3}$ and $\eta^{6}_{\alpha_1\alpha_2\alpha_3}$, but the masses extracted from them are significantly larger than those from $\eta^{1}_{\alpha_1\alpha_2\alpha_3}$ and $\eta^{2}_{\alpha_1\alpha_2\alpha_3}$. We summarize all the results in Table~\ref{tab:results}.

It is interesting to investigate the mixing of $\eta^{1}_{\alpha_1\alpha_2\alpha_3}$ and $\eta^{2}_{\alpha_1\alpha_2\alpha_3}$, since the possibly-existing physical state may have a structure much more complicated than those described by these two single currents~\cite{Chen:2008ej,Chen:2018kuu,Cui:2019roq}:
\begin{equation}
\eta^{\rm mix}_{\alpha_1\alpha_2\alpha_3}(\theta) \equiv \cos\theta~\eta^{1}_{\alpha_1\alpha_2\alpha_3} + \sin\theta~\eta^{2}_{\alpha_1\alpha_2\alpha_3} \, .
\end{equation}
However, we find that the mass minimum is arrived just at $\theta = 0^\circ$, that is $\eta^{\rm mix}_{\alpha_1\alpha_2\alpha_3}(0^\circ)  = \eta^{1}_{\alpha_1\alpha_2\alpha_3}$. Hence, this mixing does not change the extracted mass, and we shall use the results extracted from the current $\eta^{1}_{\alpha_1\alpha_2\alpha_3}$ to draw conclusions in the next section.

%
\section{Summary and Discussions}
\label{sec:summary}
%

\begin{table*}[hpt]
\begin{center}
\renewcommand{\arraystretch}{1.5}
\caption{Masses extracted from the currents $\eta^{1,2,5,6}_{\alpha_1\alpha_2\alpha_3}$.}
\begin{tabular}{c c c c c }
\hline\hline
~Currents~&~~$M_B^2~[{\rm GeV}^2]$~~&~$s_0~[{\rm GeV}^2]$~&~~Pole~[\%]~~&~~Mass~[GeV]~~
\\ \hline
$\eta^1_{\alpha_1\alpha_2\alpha_3}$&$1.32$-$1.45$&$7.2\pm1.0$&$44.9$-$53.3$&$2.33^{+0.19}_{-0.16}$
\\
$\eta^2_{\alpha_1\alpha_2\alpha_3}$&$1.33$-$1.48$&$7.6\pm1.0$&$45.1$-$54.1$&$2.45^{+0.27}_{-0.18}$
\\
$\eta^5_{\alpha_1\alpha_2\alpha_3}$&$1.46$-$1.60$&$9.6\pm1.0$&$45.1$-$53.4$&$2.72^{+0.11}_{-0.12}$
\\
$\eta^6_{\alpha_1\alpha_2\alpha_3}$&$1.45$-$1.58$&$9.4\pm1.0$&$45.2$-$53.1$&$2.67^{+0.11}_{-0.12}$
\\ \hline\hline
\end{tabular}
\label{tab:results}
\end{center}
\end{table*}

In this paper we use the method of QCD sum rules to study light tetraquark states with the exotic quantum number $J^{PC} = 3^{-+}$. We find that two quarks and two antiquarks together with at least one derivative are necessary to reach such a quantum number; besides, the quark content can be $q s \bar q \bar s$ ($q=up/down$ and $s=strange$), but can not be $s s \bar s \bar s$.

Altogether we have constructed six diquark-antidiquark interpolating currents, where the derivative can only be inside the diquark/antidiquark field, {\it i.e.},
\begin{equation}
\eta = \big[q {\overset{\leftrightarrow}{D}} s \big] \big[\bar{q}  \bar{s} \big] + \big[q s \big] \big[\bar{q} {\overset{\leftrightarrow}{D}} \bar{s} \big] \, .
\end{equation}
We use them to perform QCD sum rule analyses, and the results are summarized in Table~\ref{tab:results}. The lowest mass,
\begin{equation}
\nonumber M_{\eta_1} = 2.33^{+0.19}_{-0.16}{\rm~GeV} \, ,
\end{equation}
is extracted from the current $\eta^{1}_{\alpha_1\alpha_2\alpha_3}$, which is defined in Eq.~(\ref{def:eta1}). From its definition, we clearly see that it contains one ``good'' diquark of $s_{qs} = 1$ and one ``good'' antidiquark of $s_{\bar q \bar s} = 1$~\cite{Jaffe:2004ph}, with one of them orbitally excited:
\begin{equation}
| J^{PC} = 3^{-+} ;~s_{qs} = s_{\bar q \bar s} = 1;~l_{qs} = 1 {\rm ~or~} l_{\bar q \bar s} = 1 \rangle \, .
\label{eq:state}
\end{equation}
Since the derivative can not be between the diquark and antidiquark fields, this combination is the most stable one, phenomenologically.

In the present study we have also constructed six meson-meson interpolating currents, as defined in Eqs.~(\ref{def:xi1}-\ref{def:xi6}). Three of them have the quark combination $[\bar q q][\bar s s]$, and the derivative is between the two quark-antiquark pairs,
\begin{equation}
\xi = \big[ \bar q q \big] {\overset{\leftrightarrow}{D}} \big[ \bar s s \big] \, ;
\end{equation}
the other three have the quark combination $[\bar q s][\bar s q]$, and the derivative is inside the quark-antiquark pairs,
\begin{equation}
\xi^\prime = \big[\bar q {\overset{\leftrightarrow}{D}} s\big] \big[\bar s q\big] - \big[\bar q s\big] \big[\bar s {\overset{\leftrightarrow}{D}} q\big] \, .
\end{equation}
Hence, a special decay behavior of the $s q \bar s \bar q$ tetraquark states with $J^{PC} = 3^{-+}$ is that: a) they well decay into the the $P$-wave $(\bar q q)_{S{\rm-wave}}(\bar s s)_{S{\rm-wave}}$ final states but not into the $(\bar q q)_{S{\rm-wave}}(\bar s s)_{P{\rm-wave}}$ or $(\bar q q)_{P{\rm-wave}}(\bar s s)_{S{\rm-wave}}$ final states, and b) they well decay into the $(\bar q s)_{S{\rm-wave}}(\bar s q)_{P{\rm-wave}}$ final states but not into the $P$-wave $(\bar q s)_{S{\rm-wave}}(\bar s q)_{S{\rm-wave}}$ final states.

Especially, we use the Fierz transformation given in Eq.~(\ref{eq:fierz}) to investigate the light $s q \bar s \bar q$ tetraquark state defined in Eq.~(\ref{eq:state}). It is well coupled by the current $\eta^{1}_{\alpha_1\alpha_2\alpha_3}$, and its mass has been calculated to be $2.33^{+0.19}_{-0.16}$~GeV. Its isospin can be either $I=0$ or $I=1$, which can not be differentiated in the present study. It has a special decay behavior that: a) it well decays into the $P$-wave $\rho\phi/\omega\phi$ channel but not into the $\rho f_2(1525)/\omega f_2(1525)/\phi f_2(1270)$ channels, and b) it well decays into the $K^*(892) \bar K_2^*(1430)$ channel but not into the $P$-wave $K^*(892) \bar K^*(892)$ channel. Note that some of these features can also be derived by analysing quantum numbers of the initial and final states.

This state lies very close to the $K^*(892) \bar K_2^*(1430)$ threshold. Theoretically, it is not so easy to differentiate them, since we do not well understand the $K_2^*(1430)$ meson yet. However, experimentally, one may be able to do this, since the $K^*(892)$ and $K_2^*(1430)$ mesons are both not very narrow, {\it i.e.}, $\Gamma_{K^*(892)} = 50.3 \pm 0.8$~MeV and $\Gamma_{K_2^*(1430)} = 98.5 \pm 2.7$~MeV~\cite{pdg}. We propose to investigate the $P$-wave $\rho\phi/\omega\phi$ channel in future BESIII, Belle-II, and GlueX experiments. If there existed a narrower resonance of $J^{PC} = 3^{-+}$, it would be more likely to be a compact $s q \bar s \bar q$ tetraquark state other than a $K^*(892) \bar K_2^*(1430)$ molecular state. For completeness, in the present study we have also studied its partner state with the quark content $q q \bar q \bar q$, whose mass is extracted to be $2.27^{+0.28}_{-0.17}$~GeV.

To end this paper, we note that 
the BESIII Collaboration are possibly able to analyses some of the above decay channels simultaneously. For example, in Ref.~\cite{Ablikim:2020pgw} they performed a partial-wave analysis for the process $e^+e^- \to K^+K^-\pi^0\pi^0$. They analysed the four subprocesses $K^+(1460)K^-$, $K^+_1(1400)K^-$, $K^+_1(1270)K^-$, and $K^{*+}(892)K^{*-}(892)$, where they clearly observed the $\phi(2170)/Y(2175)$ in the former two processes but not in the latter two processes.

%
\section*{Acknowledgments}
%

We thank Wen-Biao Yan for useful discussions.
This project is supported by
the National Natural Science Foundation of China under Grants No.~11722540, No.~12005172, and No.~12075019
and
the Fundamental Research Funds for the Central Universities.

\end{document}